# Edge service resource allocation strategy based on intelligent prediction

**Yujie Wang[1], Xin Du, Xuzhao Chen and Zhihui Lu**

[1]School of computer science, Fudan University, Shanghai

Corresponding author: First A. Author (e-mail: lzh@fudan.edu.cn).

This paragraph of the first footnote will contain support information, including sponsor and financial support acknowledgment. For example, "This work was supported in part by the U.S. Department of Commerce under Grant BS123456."

**ABSTRACT** Artificial intelligence is one of the important technologies for industrial applications, but it requires a lot of computing resources and sensor data to support it. With the development of edge computing and the Internet of Things, artificial intelligence are playing an increasingly important role in the field of edge services. Therefore, how to make intelligent algorithms provide better services and the development of the Internet of Things has become an increasingly important topic. This paper focuses on the application of edge service distribution strategy, and proposes an edge service distribution strategy based on intelligent prediction, which reduces the bandwidth consumption of edge service providers and minimizes the cost of edge service providers. In addition, this article uses the real data provided by the Wangsu Technology Company and an improved long and short term memory prediction method to dynamically change the bandwidth, and achieves better optimization of resources allocation comparing with actual industrial applications.The simulation results show that our intelligent prediction can achieve good results, and the mechanism can achieve higher resource utilization.

## I. INTRODUCTION

The development of smart cities and Internet of Things (IoT) technology has been deeply ingrained in our society. It has changed the way people live and work, such as digital twins, wearable devices, smart transportation, augmented reality, etc.It also brings new challenges and opportunities to the development of edge service providers.Gartner predicts that there will be more than 15 billion Internet of Things devices connected to the enterprise infrastructure in 2029 [1]; IDC's earlier "Data Age 2025" report also pointed out that the annual global data will increase from 33ZB in 2018 to 175ZB in 2025 [2].We know that the current smart city and Internet of Things applications have created problems in the distribution of edge services.In other words, with the continuous development of the Internet of Things, edge computing has emerged. It moves the cloud computing infrastructure from the cloud far away from the user to the edge device closer to the user.Its advantage is longer-period tasks; while edge computing is more suitable for local, real-time and short-period tasks.The relationship between edge computing and cloud computing is not a substitute, but a complementary relationship.The two often need to work closely together to better meet multiple needs. As shown in Figure 1, under the heterogeneous edge cloud architecture, there are not only centralized cloud service platforms, but also a large number of heterogeneous edge nodes and edge devices.Through the unified management of the edge-cloud hybrid system, the application and service support requested by users are realized. Because the edge node is deployed at the edge layer, it usually only has a resource pool composed of a few servers or micro data centers [3]. However, since various devices of the terminal are connected to the system platform through the edge layer, so generally speaking,the pressure of resource shortage in the edge layer is relatively large, and a centralized cloud platform is needed to complete the cooperative work with the platform on the edge side, so as to provide sufficient storage resources and computing resources for the edge cloud hybrid system.In real life, the allocation of edge service resources in scenarios such as smart transportation, smart security, large parking lots or airports has become an important problem.

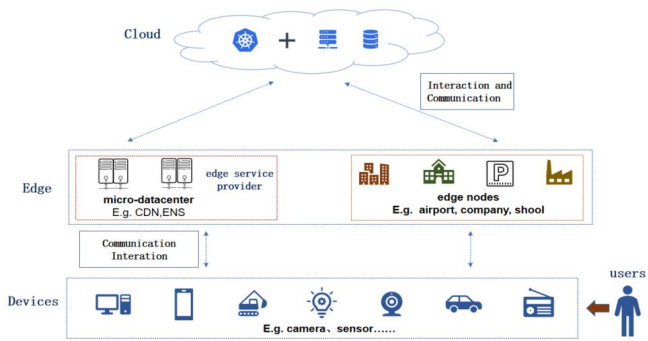

Figure 1: the heterogeneous edge cloud architecture

In order to better solve the problem of edge service distribution, the solution we adopted is to use predicted bandwidth traffic to provide a reference for service providers. Accurately predicting traffic can reduce the number of migrations of edge service resources such as bandwidth, and reduce transmission costs and energy costs. Most methods of predicting early traffic use statistical analysis. However, statistical methods cannot comprehensively consider the overall scene factors. Although the neural network prediction algorithm was proposed a few years ago, the effect is not good due to the problem of the disappearance of the gradient in the back propagation. In recent years, nonlinear activation functions have been widely used in such scenarios. And we will adopt intelligent algorithms to achieve this goal.

Although artificial intelligence algorithms have been applied in many fields, such as natural language processing [4] and speech recognition [5], most machine learning methods require a large amount of data and training time to train an accurate models. Therefore, the application has many limitations. The combination of edge computing and artificial intelligence can improve the weaknesses of artificial intelligence. Its application can be simply divided into two parts. The first part is the improvement of the flow, through simple operation to filter out the useless data in advance, reducing the back haul transmission. Although edge computing can improve computing efficiency, it still takes a lot of training time when the amount of data is large. Therefore, it is difficult to apply to low-latency environments such as Internet of Vehicle (IoV). It has become a trend to combine other mechanisms to realize the distribution of edge services. The meta-heuristic algorithm can find a good solution in a limited time. For edge service distribution scenarios with real-time requirements, they are potential solutions, such as simulated annealing [6], tabu search [7], genetic algorithm [8], ant colony optimization [9], particle swarm optimization [10] ,etc. This paper proposes a new edge service resource allocation strategy based on intelligent prediction. It not only uses the improved LSTM intelligent prediction algorithm to predict the service provider bandwidth of real data, but also uses the prediction results to provide a scheme for the allocation

of edge service resources. In summary, the research contributions we provide in this article are as follows:

1. Considering the real-time bandwidth problem of low transmission delay and low power consumption, we use an improved long short-term memory (LSTM) to predict the bandwidth flow and the usage of various server resources.

2. Through the bandwidth prediction results obtained by the intelligent prediction algorithm provide a new edge service resource allocation scheme for the edge service providers.

3. Based on the real data, the prediction of bandwidth resources and the verification of resource allocation scheme are carried out, and the conclusion is drawn that our prediction method and allocation scheme can be used effectively.

The rest of this article is organized as follows. The second section introduces the related work of edge service resource allocation and traffic prediction. The third section gives the proposed specific scenario model and method, including resource allocation and traffic prediction mechanism, and the fourth section gives the simulation results and discussion. The fifth part summarizes this paper and prospects the future work.

## II. RELATED WORK

Researcher Farhadi V et al. [11] from the United States conducted joint research on edge service deployment and task scheduling, but the targeted applications are applications with large data volumes and large transmission volumes. Dehghan M et al. [12] aimed at minimizing application access time for congestion-sensitive applications, considering which content is deployed on which nodes and how to access it. He T et al. [13] conducted a joint study on service deployment and task scheduling under the condition that storage resources can be shared, and computing and network resources cannot be shared. The results prove that it can serve more requests than independent studies and improve resource utilization. In addition, the research group of Professor Weisong Shi of Wayne State University has sorted out the development process of edge computing in the literature [14] around "where does edge computing come from, what is its current situation, and where it is going", and summarized it as the three stages: technology reserve period, rapid growth period and steady development period and enumerate typical events in different stages, and put forward the problems that need to be solved urgently in the future development of edge computing.

In domestic research, the research group of Professor Deng Shuiguang of Zhejiang University proposed that the greedy redundancy deployment algorithm SAA-RS [15], which approximates the average of samples, can be used when deploying services. It through long-term observation of user requests and use the method of sampling mean approximation to solve the uncertainty of the user's request for access.The research group also modeled services as a combination of microservice instances, took the service deployment problem as the deployment problem of microservice instances, and studied the optimization of service deployment under the condition of limited edge node resources [16].In order to improve the IoT data management scheme of blockchain and edge computing [17 ], they designed a distributed data storage scheme to improve the storage efficiency of the system; they also proposed a blockchain-based active access control mechanism , Designed a built-in encryption scheme to ensure data security and privacy.Professor Tan Haisheng's research group of the University of Science and Technology of China fully considered the deadline of the task when scheduling the task in the literature [18],and proposed the Dedas algorithm to make as many tasks as possible to complete on time under the condition that the task has a deadline and the edge nodes have bandwidth constraints.Because this algorithm adjusts the order of tasks in the waiting queue on the edge node to achieve more tasks to be completed, it is not friendly enough for some applications that take up a lot of bandwidth.The research group of Professor Wang Shangguang of Beijing University of Posts and Telecommunications makes full use of the lightweight, fast-starting, and fast-migrating characteristics of microservice technology to dynamically deploy applications to solve the delay problem caused by users' rapid movement [19].This document models the collaboration of microservices as a shortest path problem with a forward-looking time window and uses a static offline algorithm to obtain an optimal collaborative migration deployment plan. At the same time, it also designs a dynamic online collaboration algorithm based on reinforcement learning to achieve the approximate optimal effect.The research group also studied the service deployment and task scheduling problems in edge computing. Based on the Gibbs sampling method, the strategy of service deployment was continuously updated and the task scheduling algorithm was proposed based on the idea of "water injection", which effectively reduced the service response time and the number of requests that are forwarded to the cloud,but the research on service deployment and task scheduling is decoupled, and the calculation time for service deployment is long [20]. In order to better summarize and prospect the research field, in the literature [21], the team also introduced the concept of edge computing, typical application scenarios, research status and key technologies, etc., and proposed the development of edge computing is in the initial stage and there are still many problems to be solved in practical applications, including optimization of edge computing performance, security, interoperability, and intelligent edge operation management services.The research group of Professor Li Dongsheng of the National University of Defense Technology introduced the system architecture characteristics of edge computing applications in the literature [22], analyzed the causes of long-tail delay, and classified the main theories and methods of network delay measurement, and summarized the optimization techniques for long tail delay.Finally, the idea of the operating environment online and the challenges it faces are put forward.The research group of Professor Gao Ling from Northwestern University proposed a deep model classification task scheduling strategy for edge devices [23]. The strategy through collaborative mobile devices and edge server, make full use of the convenience of smart mobile terminals and the edge server powerful computing ability, considering the complexity of the classification task and user expectations and complete the dynamic deployment of deep models in mobile devices and edge servers,so as to improve the task execution efficiency and reduce the inference overhead of the deep learning model.

In order to apply the edge cloud architecture to actual industrial cases, many domestic scientists have conducted in-depth and extensive investigations from various angles.The research group of Professor Zheng Zibin and Huang Huawei of Sun Yat-sen University not only proposed an online algorithm for drone swarms' computing offloading and multi-hop routing scheduling jointly optimized in the edge cloud environment to meet the requirements of the drone swarm for delay and computing power in the edge environment [24], a novel management architecture that provides NFV services in distributed edge computing is also proposed, which can provide active failure recovery for NFV-enabled distributed edge computing [25].Chen Xu's team at Sun Yat-Sen University proposed an efficient online cloud edge resource configuration framework based on time-lag-aware Lyapunov optimization technology [26], which can provide online greedy decision-making of edge cloud resource allocation for heterogeneous IoT applications. They also proposed an online learning algorithm based on Thompson-sampling to explore the characteristics of the dynamic edge cloud environment, and further help mobile users make adaptive service location decisions [27].The team of Professor Xu Xiaolong from Nanjing University of Information Science and Technology proposed a service diversion method based on deep reinforcement learning for the Internet of Vehicles in edge computing [28], so that the vehicle can effectively adapt to the situation of excessive service requests, thereby improving the user's quality of service (QoS).Professor Huang Jiwei's research group from China University of Petroleum proposed a simulation-based QoS-aware dynamic service selection method for mobile edge computing systems [30], gave a stochastic system model and mathematical analysis, and applied target softening to the original problem. Through the optimization technology design service selection algorithm, the effectiveness of the method is verified through simulation experiments.The team of Professor Li Jianzhong of Harbin Institute of Technology optimized the WPT time allocation and calculation scheduling of mobile devices in the WP-MEC network to maximize the calculation completion rate of the WP-MEC network, and proposed an approximate algorithm [31]. The team of Professor Xu Zichuan of Dalian University of Technology studied the problem of service caching in mobile edge networks, proposed an integer linear programming (ILP) and random rounding algorithm, and designed a distributed and stable game theory mechanism. By allowing network service providers to cooperate in the service cache, the social costs of all participants are significantly reduced [32].The team of Professor Song Lingyang of Peking University proposed a multi-layer data flow processing system EdgeFlow, which comprehensively utilizes the cloud computing center (CC) layer, and edge mobile computing realizes efficient data processing at the bottom of the middle layer (MEC) server and edge device (EDs) layer[33]. Not only that, they also modeled the interaction between data service operators and authorized data service users as an equilibrium problem with equilibrium constraints, and used the alternating direction method of multipliers as a large-scale optimization tool to solve them, thereby optimizing DSOs The resource pricing and the number of resources ADSSs need to purchase. At the same time, a model based on many-to-many matching is proposed to achieve effective resource allocation in IoT fog computing supported by NFV [34]. The team of Professor Ren Ju from Central South University studied the cache deployment problem of a large-scale WiFi system with 8,000 APs serving more than 40,000 active users, and designed a TEG (Traffic Weighted Greedy) algorithm to solve the problem of maximizing long-term cache gain and to maximize long-term cache benefits. In addition, the research group of Professor Liu Zhezhe of Peking University introduced RetroI2V-a new type of vehicle communication and network system in the literature. It can recognize traditional road signs and transmit additional dynamic information to vehicles. RetroI2V can use the reflective coating of road signs to establish visible light backscatter communication (VLBC) to further coordinate multiple concurrency VLBC session between road signs and approaching vehicles.

In summary, in the field of edge service distribution, although both domestic and foreign teams have made important contributions, these distribution schemes can no longer meet the huge demand for edge services. It is necessary to consider the impact of multiple edge service resources on the entire network, as well as the increasing problem of large-scale resource allocation. The introduction of intelligent algorithms into this field has become a concern of more and more research.

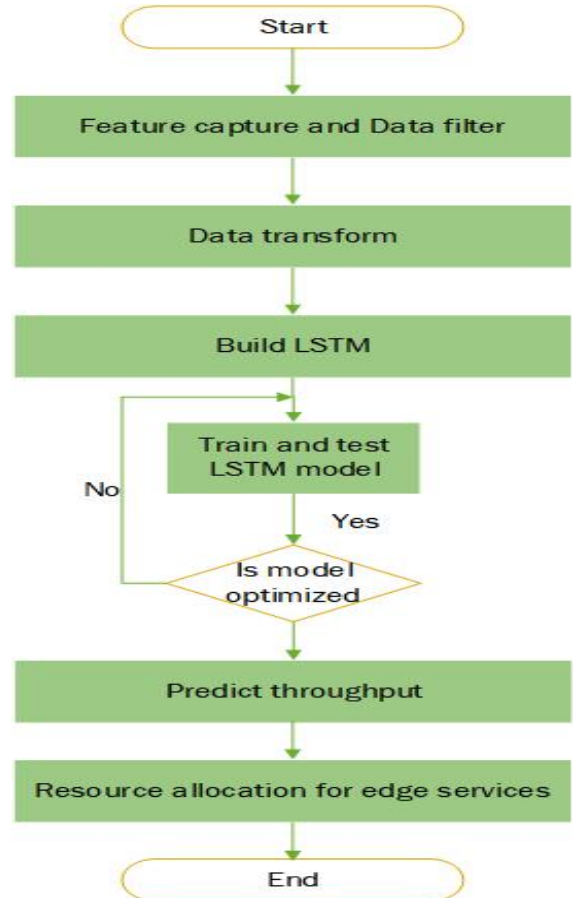

Figure 2: a simple flow chart of the method

## III. METHOD

This section mainly introduces and explores three parts in detail from the allocation process of edge service resources, using the improved LSTM intelligent algorithm to predict traffic and the resource allocation plan based on the prediction result of the intelligent algorithm. A series of narratives is made for these three parts divided into three subsections.

1. The allocation process of edge service resources

We use an improved LSTM to predict the bandwidth traffic of edge service providers, and use the prediction data as an important basis for edge service resource allocation. Then, use the traffic usage of the edge server in each period to specifically allocate various resources. Our goal is to minimize the use of various resources on the server and the number of migrations of edge services to reduce the cost of edge service providers.First, the edge service provider will randomly connect to the edge server and send a request. The equipment service provider will preprocess the collected traffic for a period of time through the edge server, including feature extraction, data filtering and data conversion. Then build an improved LSTM model to learn the daily resource usage of each service provider. If the loss function meets the termination condition, the model trained for the service provider can be used to predict traffic, otherwise, it will continue training.Through the prediction of each edge service provider, we can make the edge service provider with less demand for edge service resources shut down or let the overloaded edge server offload the bandwidth traffic to other edge servers. Therefore, resource allocation will be more accurate. Finally, based on this predicted solution, the overall server resource allocation is performed, and the resource allocation situation is repeatedly selected and updated until the termination condition is met. Through the above-mentioned processes and mechanisms, we can dynamically allocate resources for edge servers, minimizing the overall cost of all edge service providers. The detailed flow chart of the process is shown in Figure 2 (a simple flow chart is drawn here).

2. Improved LSTM intelligent algorithm for traffic prediction

In the solution for forecasting the bandwidth resources of edge service providers, we use the hour as the basic unit and make a forecast every hour. By trying to use the improved LSTM intelligent algorithm to predict the bandwidth traffic of future service providers, 80% of the entire data used for prediction is used as the training set, and 20% of the entire data is used as the predicted value. It is especially worth noting that the bandwidth traffic of a service provider will be affected by various emergencies, such as the update of new versions of mobile games, the live broadcast of large-scale events, etc. Considering the emergence of sudden traffic, the design time prediction sequence length Is a shorter time length. In other words, the essence of this method is to predict the time-varying sequence, so it is a good choice to realize it through the method of intelligent algorithm. In addition, it is also particularly important to evaluate the accuracy of the predicted bandwidth resource usage based on the size of the bandwidth resource used by the actual edge service provider. We use the mean square error to calculate the loss here. The loss function can be used as a standard for evaluating the accuracy of the prediction. The smaller the value of the loss function, the more accurate the prediction.

In order to avoid that the value of some data is too large or too small to affect the prediction results, we filtered out part of the abnormal traffic that is visible to the naked eye in the process of data preprocessing. After filtering out the data, we convert the data into a data pattern that conforms to the Tensor process. This article uses LSTM to build a model, which is divided into three layers, including input layer, hidden layer and output layer. The loss function uses Huber; the optimizer uses Adam. The activation function uses Relu to reduce the problem of gradient disappearance, minimize the prediction loss value after training, and then improve the prediction ability. The running process of LSTM in our scenario is shown in Figure 3.

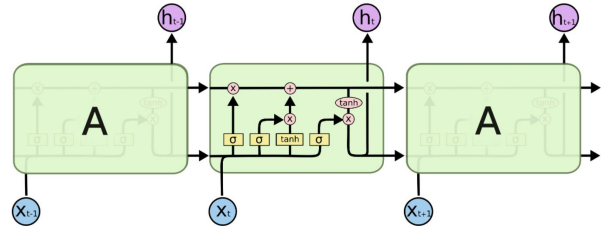

Figure3: the running process of LSTM

Through the above process, we have obtained the predicted value of the network bandwidth of the edge service provider in each time period.

3. Edge service resource allocation plan based on the prediction results of the intelligent algorithm.

Equipment providers can provide a variety of optional resource allocation schemes for edge service providers. Taking containers as an example, the container configurations that can be selected by edge service providers include the number of CPU cores, memory size, disk size, etc., cluster configuration includes the configuration of the container and the number of containers.It can be seen that the allocation of edge service resources is discrete and limited. After research on different edge service providers, it is found that most edge service providers have the same resource usage, that is, general-purpose resource allocation schemes can be constructed for the same type of edge service providers. This idea of bandwidth allocation by predicting the allocation of bandwidth resources provides a solid theoretical foundation and valuable practical exploration for industrial applications of edge services, and promotes the application of basic research results.

## IV. EXPERIMENT AND RESULTS

1. Experimental environment

Our environment is the Linux operating system, using the Python programming language, and Anaconda as the Python execution environment and package management. The installed TensorFlow kit is the 1.0.1 GPU version. Since TensorFlow supports a distributed computing architecture, it can be deployed on multiple computing devices. In terms of computing resource allocation, using edge AI to preprocess data can not only reduce the transmission of useless data in the backhaul, but also filter out unnecessary data. Cloud servers are used to process more complex calculations, such as neural networks and GARAA, so that deep learning has better performance. If the amount of data is large and the model complexity is high, multiple machines can be used in the cloud server to achieve acceleration and solve real-time requirements.

2. Data introduction

The edge service provider meets the user's delay requirements by deploying the provided services on the edge nodes. Wangsu Technology Company provides a large number of logs of edge hardware resources leased by service providers. Analysis shows that 380 service providers need to use the edge nodes provided by Wangsu every day. The data can also be analyzed to obtain the resources on the edge nodes used by each service provider in real time, which provides the conditions for the resource configuration and the coordinated scheduling of services studied in this paper.

3. Experimental results

Based on the current 14-day data from 2020.12.12 to 2020.12.23, a total of 665 customer data are counted. Figure 4 is a broken line graph based on the 14-day bandwidth change of a certain edge service provider over time. It can be seen that the bandwidth has a relatively obvious relationship change over time.

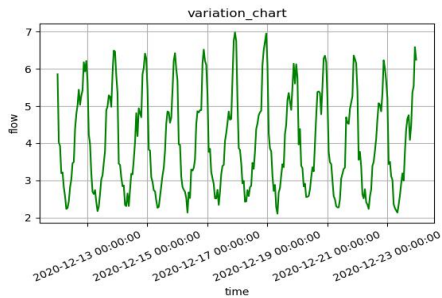

Figure 4: the 14-day bandwidth change of a certain edge service provide

We also made a quantitative comparison of the loss error of the prediction through the mean square error, the root mean square error and the average absolute error. Their specific calculation methods are as follows:

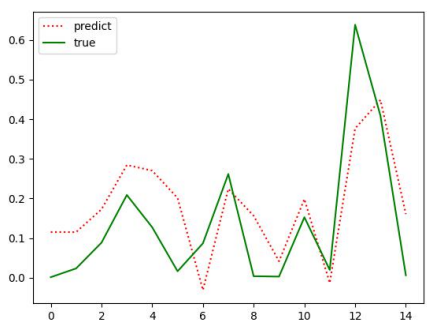

Figure 5: The prediction of intelligent methods

The final prediction result is that the mean square error is equal to 0.000233, the root mean square error is equal to 0.015278, and the average absolute error is equal to 0.011488.

## V. CONCLUSION

Artificial intelligence has led the rapid development of many industries, but many applications are based on high-speed transmission and powerful computing resources. In the future, the development of a large number of edge services will meet people's increasing material needs, but it also brings some challenges, the most important of which is the resource allocation of edge services. Therefore, based on the current development trend, this article proposes a new architecture that combines edge computing and cloud computing to effectively predict the bandwidth resources used by each service provider, and then effectively promote the allocation of various resources for edge services. In future work, we will focus on improving the accuracy of throughput prediction and improving the real-time performance of prediction methods.

## REFERENCES

[1] https://www.gartner.com/

[2] https://www.idc.com/analysts

[3] Du X, Tang S, Lu Z, et al. A Novel Data Placement Strategy for Data-Sharing Scientific Workflows in Heterogeneous Edge-Cloud Computing Environments[C]//2020 IEEE International Conference on Web Services (ICWS). IEEE, 2020: 498-507.

[4] T. Young, D. Hazarika, S. Poria, and E. Cambria, "Recent trends in deep learning based natural language processing," IEEE Comput. Intell. Mag., vol. 13, no. 3, pp. 55–75, Aug. 2018.

[5] W. Xiong, L. Wu, F. Alleva, J. Droppo, X. Huang, and A. Stolcke, "The Microsoft 2017 conversational speech recognition system," in Proc. IEEE Int. Conf. Acoust., Speech Signal Process., Apr. 2018, pp. 5934–5938.

[6] L. Wei, Z. Zhang, D. Zhang, and S. C. Leung, "A simulated annealing algorithm for the capacitated vehicle routing problem with two-dimensional loading constraints," Eur. J. Oper. Res., vol. 265, no. 3, pp. 843–859, Mar. 2018.

[7] G. L. A. M´endez, P. E. J. G´omez, and V. A. Terr´e, "Application of Tabu search based algorithms for symbol detection in L-MIMO systems," in Proc. IEEE Colombian Conf. Commun. Comput.,

Cartagena, Colombia, USA, Aug. 2017, pp. 1‑6.

[8] G. Oliveri et al., "Codesign of unconventional array architectures and antenna elements for 5G base stations," IEEE Trans. Antennas Propag., vol. 65, no. 12, pp. 6752‑6767, Dec. 2017.

[9] B. Chen, J. Zhang, Q. Zhu, X. Wang, and M. Gao, "Energy-efficient traffic grooming in 5G C-RAN enabled flexible bandwidth optical networks," in Proc. Asia Commun. Photon. Conf., Nov. 2017, pp. M3B‑2.

[10] T. Shahjabi and K. V. Babu, "Cooperative scheme for wireless energy harvesting and spectrum sharing in cognitive 5G networks," in Proc. Int. Conf. Innovations Elect., Electron., Instrum. Media Technol., Feb. 2017, pp. 57‑61.

[11] Farhadi V, Mehmeti F, He T, et al. Service placement and request scheduling for data-intensive applications in edge clouds[C]//IEEE INFOCOM 2019-IEEE Conference on Computer Communications. IEEE, 2019: 1279-1287.

[12] Dehghan M, Jiang B, Seetharam A, et al. On the complexity of optimal request routing and content caching in heterogeneous cache networks[J]. IEEE/ACM Transactions on Networking, 2016, 25(3): 1635-1648.

[13] He T, Khamfroush H, Wang S, et al. It's hard to share: Joint service placement and request scheduling in edge clouds with sharable and non-sharable resources[C]//2018 IEEE 38th International Conference on Distributed Computing Systems (ICDCS). IEEE, 2018: 365-375.

[14] Cai, W., Du, X., & Xu, J. (2019). A personalized QoS prediction method for web services via blockchain-based matrix factorization. Sensors, 19(12), 2749.

[15] Zhao H, Deng S, Liu Z, et al. Distributed Redundancy Scheduling for Microservice-based Applications at the Edge[J]. IEEE Transactions on Services Computing, 2020.

[16] Deng S, Xiang Z, Taheri J, et al. Optimal application deployment in resource constrained distributed edges[J]. IEEE Transactions on Mobile Computing, 2020.

[17] 程冠杰, 黄净杰, 邓水光. 基于区块链与边缘计算的物联网数据管理[J]. 物联网学报, 4(2): 1-9.

[18] Meng J, Tan H, Xu C, et al. Dedas: Online task dispatching and scheduling with bandwidth constraint in edge computing[C]//IEEE INFOCOM 2019-IEEE Conference on Computer Communications. IEEE, 2019: 2287-2295.

[19] Wang S, Guo Y, Zhang N, et al. Delay-aware microservice coordination in mobile edge computing: A reinforcement learning approach[J]. IEEE Transactions on Mobile Computing, 2019.

[20] Ma X, Zhou A, Zhang S, et al. Cooperative service caching and workload scheduling in mobile edge computing[J]. arXiv preprint arXiv:2002.01358, 2020.

[21] Du, X., Xu, J., Cai, W., Zhu, C., & Chen, Y. (2019). Oprc: An online personalized reputation calculation model in service-oriented computing environments. IEEE Access, 7, 87760-87768.

[22] Du, X., & Cai, W. (2018, November). Simulating a basketball game with hdp-based models and forecasting the outcome. In 2018 7th International Conference on Digital Home (ICDH) (pp. 193-199). IEEE.

[23] 任杰, 高岭, 于佳龙,等. 面向边缘设备的高能效深度学习任务调度策略[J]. 计算机学报, 2020, 043(003):440-452.

[24] Liu B, Zhang W, Chen W, et al. Online Computation Offloading and Traffic Routing for UAV Swarms in Edge-Cloud Computing[J]. IEEE Transactions on Vehicular Technology, 2020, 69(8): 8777-8791.

[25] Huang H, Guo S. Proactive failure recovery for NFV in distributed edge computing[J]. IEEE Communications Magazine, 2019, 57(5): 131-137.

[26] Zhou Z, Yu S, Chen W, et al. CE-IoT: Cost-Effective Cloud-Edge Resource Provisioning for Heterogeneous IoT Applications[J]. IEEE Internet of Things Journal, 2020, 7(9): 8600-8614.

[27] Ouyang T, Li R, Chen X, et al. Adaptive user-managed service placement for mobile edge computing: An online learning approach[C]//IEEE INFOCOM 2019-IEEE Conference on Computer Communications. IEEE, 2019: 1468-1476.

[28] Xu X, Shen B, Ding S, et al. Service offloading with deep Q-network for digital twinning empowered Internet of Vehicles in edge computing[J]. IEEE Transactions on Industrial Informatics, 2020.

[29] Jiwei Huang, Yihan Lan, Minfeng Xu: A Simulation-Based Approach of QoS-Aware Service Selection in Mobile Edge Computing. Wirel. Commun. Mob. Comput. 2018: 5485461:1-5485461:10 (2018).

[30] Zhu T, Li J, Cai Z, et al. Computation scheduling for wireless powered mobile edge computing networks[C]//IEEE INFOCOM 2020-IEEE Conference on Computer Communications. IEEE, 2020: 596-605.

[31] Xu Z, Zhou L, Chau S C K, et al. Collaborate or

separate? Distributed service caching in mobile edge clouds[C]//IEEE INFOCOM 2020-IEEE Conference on Computer Communications. IEEE, 2020: 2066-2075.

[32] Wang P, Yao C, Zheng Z, et al. Joint task assignment, transmission, and computing resource allocation in multilayer mobile edge computing systems[J]. IEEE Internet of Things Journal, 2018, 6(2): 2872-2884.

[33] Raveendran N, Zhang H, Song L, et al. Pricing and Resource Allocation Optimization for IoT Fog Computing and NFV: An EPEC and Matching Based Perspective[J]. IEEE Transactions on Mobile Computing, 2020.

[34] Lyu F, Ren J, Cheng N, et al. Demystifying traffic statistics for edge cache deployment in large-scale WiFi system[C]//2019 IEEE 39th International Conference on Distributed Computing Systems (ICDCS). IEEE, 2019: 965-975.